\documentclass{ws-mpla}

\begin{document}

\markboth{L. Ya. Glozman and R. F. Wagenbrunn}
{}

\catchline{}{}{}{}{}

\title{CHIRALLY SYMMETRIC BUT CONFINED HADRONS AT FINITE DENSITY}

\author{\footnotesize L. Ya. GLOZMAN}

\address{Institute for
 Physics, Theoretical Physics branch, University of Graz\\
 Universit\"atsplatz 5, A-8010 Graz, Austria\\
leonid.glozman@uni-graz.at}

\author{R. F. WAGENBRUNN}

\address{Institute for
 Physics, Theoretical Physics branch, University of Graz\\
 Universit\"atsplatz 5, A-8010 Graz, Austria
}

\maketitle

\pub{Received (Day Month Year)}{Revised (Day Month Year)}

\begin{abstract}
At a critical finite chemical potential and low temperature 
QCD undergoes the chiral restoration phase transition. The
folklore tradition is that simultaneously hadrons are
deconfined and there appears the quark matter. We demonstrate
that it is possible to have confined but chirally symmetric
hadrons at a finite chemical potential and hence beyond the
chiral restoration point at a finite chemical potential and low
temperature there could exist a chirally symmetric matter
consisting of chirally symmetric but confined hadrons.
If it does happen in QCD, then the QCD phase diagram should
be reconsidered with obvious implications for heavy ion programs
and astrophysics.

\end{abstract}

\ccode{PACS Nos.:11.30.Rd, 25.75.Nq, 12.38.Aw }

\medskip
{\bf 1. Introduction.}
It is generally believed that chiral and deconfinement phase
transitions in QCD coincide and hence beyond the semi-circle in 
the $T-\mu$ plane one obtains a deconfining
and chirally restored matter. At zero temperature and density
the 't Hooft anomaly matching conditions require that in the
confining phase  chiral symmetry must be broken in the vacuum.
However, there is no such a restriction at a finite chemical
potential. Typically  models that give us 
information about the
phase structure in QCD are of the Nambu and Jona-Lasinio type.
These models are not confining, however, and consequently there
is no  basis to conclude that above the chiral restoration
point at finite chemical potential one obtains a chirally symmetric
and deconfining quark matter.

Recently McLerran and Pisarski presented qualitative large $N_c$ arguments
showing that at reasonably large chemical potential and low temperature
there might exist a confining but chirally symmetric phase  \cite{pisarski}.
This suggestion is in conflict with the naive intuition that once
the hadrons are confined chiral symmetry should be broken. Here we
demonstrate that it is not so and that it is possible to have confined
but chirally symmetric hadrons at finite density and low temperature \cite{GW}.

There exists only one known manifestly chirally-symmetric and confining
model in four dimensions that is solvable \cite{Orsay}.
This model
can be considered as a generalization of the 1+1 dimensional
't Hooft  model, that is QCD in the large $N_c$ limit.
Once the gauge is properly chosen in 1+1 dimensions 
the Coulomb interaction becomes a linear confining
potential.  Then  this
potential properly represents gluonic degrees of freedom in 1+1
dimensions. 
It is postulated that there
exists a linear confining potential of the Coulomb type in four dimensions
either.
 This model represents
a simplification of  large $N_c$ QCD in four dimensions.  
The model is exactly solvable. 

\medskip
{\bf 2. Chiral symmetry breaking in the vacuum.}
Consider first chiral symmetry breaking in the vacuum.
The Dirac operator for the dressed quark is
\begin{equation}
D(p_0,\vec{p})= i S^{-1}(p_0,\vec{p}) = D_0(p_0,\vec{p})-\Sigma(p_0,\vec{p}),
\label{SAB}
\end{equation}

\noindent
where  $D_0$ is the bare Dirac operator. 
Parametrising the self-energy operator in the form

\begin{equation}
\Sigma(\vec p) =A_p +(\vec{\gamma}\hat{\vec{p}})[B_p-p],
\label{SE} 
\end{equation}

\noindent
where functions $A_p$ and $B_p$ are yet to be found, the
Schwinger-Dyson equation for the self-energy operator in 
the rainbow approximation,which is valid in the large $N_c$ limit
for the instantaneous interaction, 
 is reduced to the nonlinear gap equation for the chiral angle
 $\varphi_p$,
 
 \begin{equation}
 A_p \cos \varphi_p - B_p \sin \varphi_p = 0,
 \label{gap}
 \end{equation}
 
\noindent
where
 
\begin{eqnarray}
A_p & = & \frac{1}{2}\int\frac{d^3k}{(2\pi)^3}V
(\vec{p}-\vec{k})\sin \varphi_k,\quad  \\
B_p & = & p+\frac{1}{2}\int \frac{d^3k}{(2\pi)^3}\;(\hat{\vec{p}}
\hat{\vec{k}})V(\vec{p}-\vec{k})\cos \varphi_k. 
\label{AB} 
\end{eqnarray}  

\noindent
Here $V(\vec p)$ is the Fourier transform of the linear confining potential
with the string tension $\sigma$ with a proper infrared regularisation. 
The functions $A_p,B_p,$  i.e. the quark self-energy,
  are divergent in the
 infrared limit, which implies that the single quark cannot be observed
 and the system is confined. However, the infrared divergence 
 cancels exactly in 
 the gap equation 
so this equation can be solved directly in the infrared limit. The
chiral symmetry breaking is signalled by the nonzero quark condensate
and by the dynamical momentum-dependent mass of quarks

\begin{equation}
\langle\bar{q}q\rangle=-\frac{N_C}{\pi^2}\int^{\infty}_0 dp\;p^2\sin\varphi_p,
~~~~~~~
M(p) = p \tan \varphi_p.
\label{dyna}
\end{equation} 

\noindent
The dynamical mass is finite at small momenta and vanishes
at large momenta.
Both these quantities were first obtained in ref. \cite{Adler:1984ri}
and repetedly reconfirmed in all subsequent works on this model.
The numerical value of the quark condensate is 
$\langle\bar{q}q\rangle=(-0.231\sqrt{\sigma})^3$. 

\medskip 
{\bf 3. Chiral symmetry restoration and  meson spectra.}
Now we are in position to include into this model a finite quark
chemical potential $\mu$ at zero temperature. Denoting the Fermi
momentum of quarks as $p_f$ one has to remove from the integration
 in the gap  equation all
quark momenta below $p_f$ since they are Pauli-blocked. This is similar
to what is done within the NJL type models or within the 't Hooft model
\cite{Thies}. At the critical quark chemical potential one observes a
chiral restoration phase transition, as depicted in Fig. 1.

\begin{figure}
\includegraphics[width=0.6\hsize,clip=]{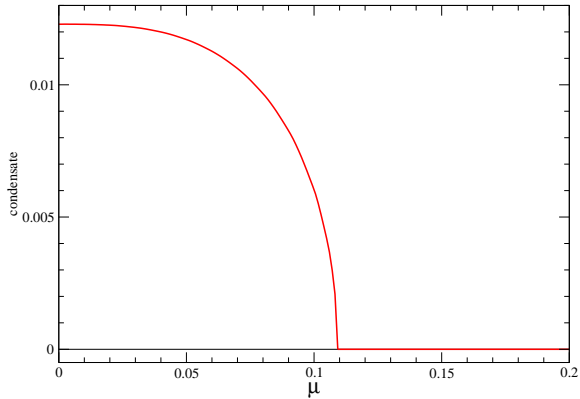}
\caption{Quark condensate in units of $\sigma^{3/2}$
as a function of the chemical potential, which is in units of $\sqrt \sigma$}
\end{figure}

What
crucially distinguishes this model from the NJL model is that the system
is still confined, even in the chirally restored phase. This is because
the self-energy integral $B_p$ is still infrared-divergent, even when
the chiral angle $\varphi_p$ and dynamical mass $M(p)$ identically
vanish. Hence the single quark 
 is removed from the spectrum at any chemical potential.
 
 To demonstrate this explicitly one has to solve the Bethe-Salpeter
 equation for the quark-antiquark bound states
 applying the quark Green function obtained from the gap
 equation. The infrared divergence of the single-quark Green 
 function
cancels exactly in the color-singlet quark-antiquark system
\cite{WG2} and the bound state mesons are finite and well defined
quantities. The spectrum below the critical chemical potential
is similar to the one previously obtained in the vacuum \cite{WG2}.
The spectrum exhibits 
approximate restoration
of the chiral symmetry in excited hadrons, for details we refer to \cite{WG2}
and for a review to ref. \cite{GPR}.

The spectrum above the critical chemical potential, i.e. for 
$\mu = 0.2\sqrt{\sigma}$, is shown
in Fig. 2. This spectrum is qualitatively different. 
All the states are in {\it exact} chiral multiplets. 
Though the chiral symmetry
is manifestly restored in the vacuum, one observes
{\it finite-energy well defined hadrons}. Obviously the mass generation
mechanism in these hadrons has nothing to do with the chiral symmetry breaking
in the vacuum and is not related with the quark condensate. The mass
generation mechanism for these chirally symmetric hadrons is similar to
the high-lying states in the chiral symmetry broken phase. 

\begin{figure*}
\mbox{
\includegraphics[width=0.16\hsize]{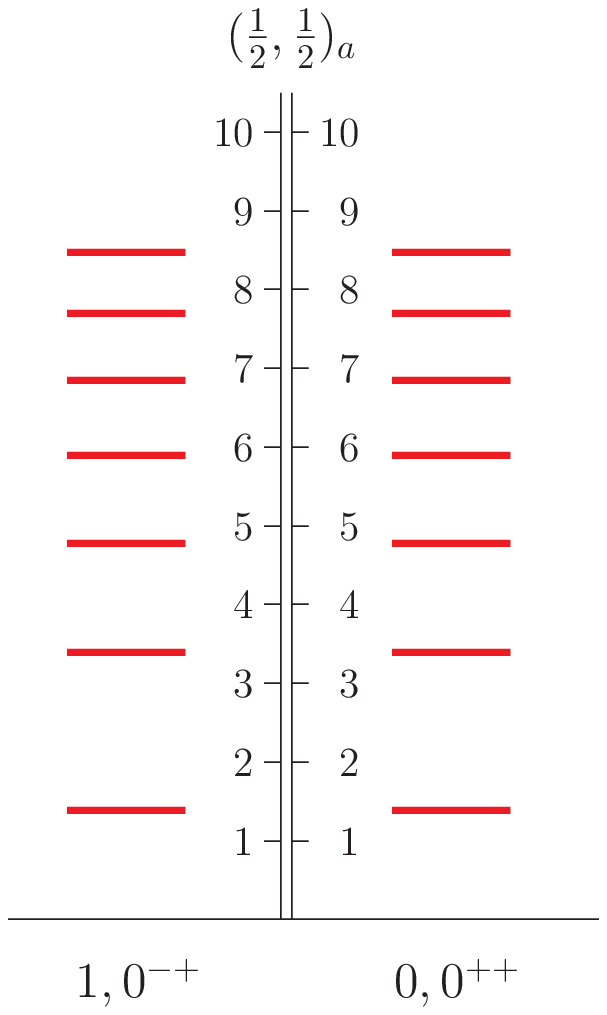}\,%
\includegraphics[width=0.16\hsize]{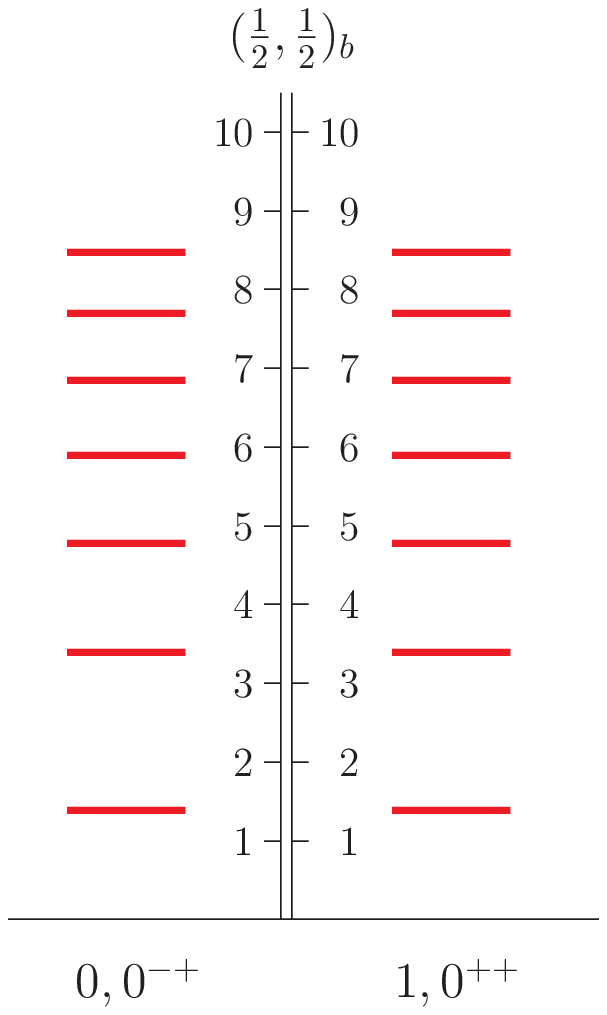}\,%
\includegraphics[width=0.16\hsize]{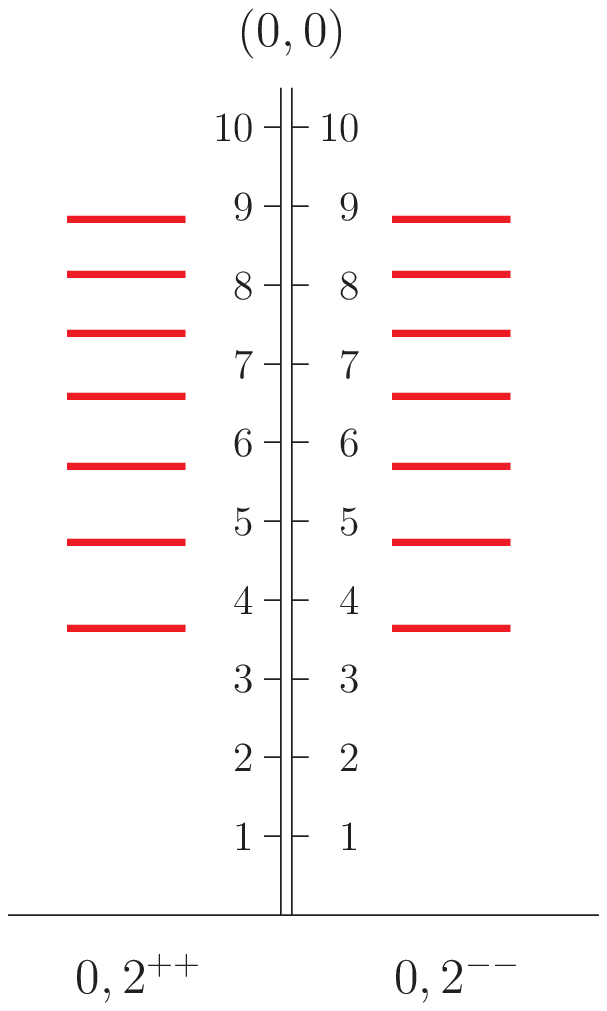}\,%
\includegraphics[width=0.16\hsize]{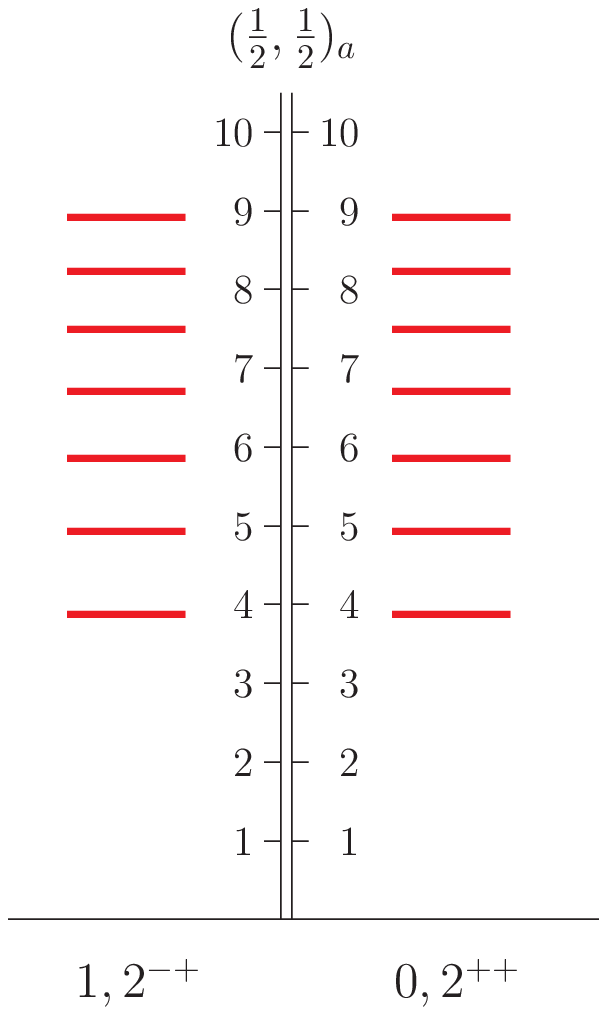}\,%
\includegraphics[width=0.16\hsize]{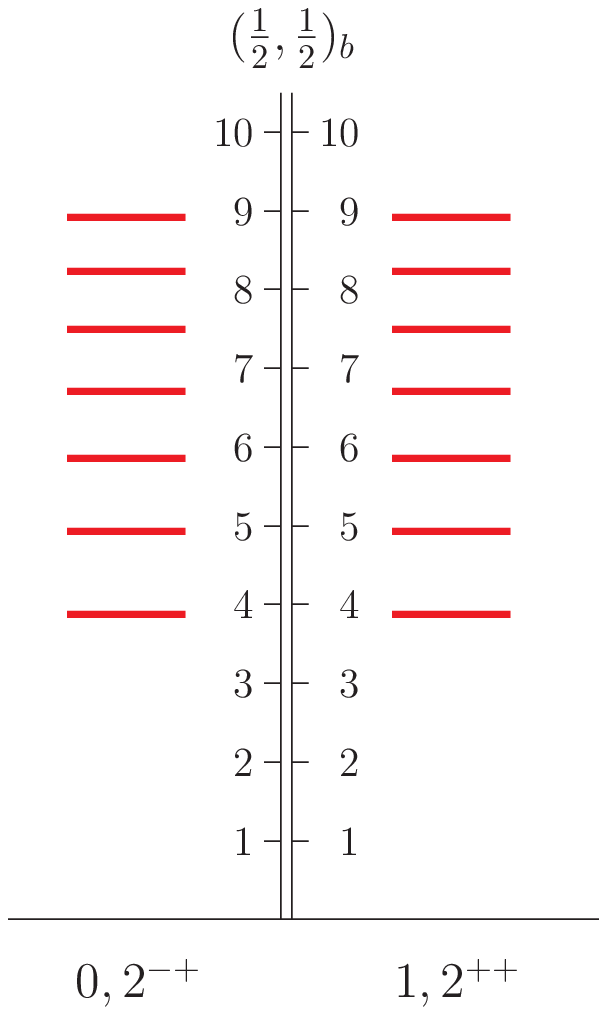}\,%
\includegraphics[width=0.16\hsize]{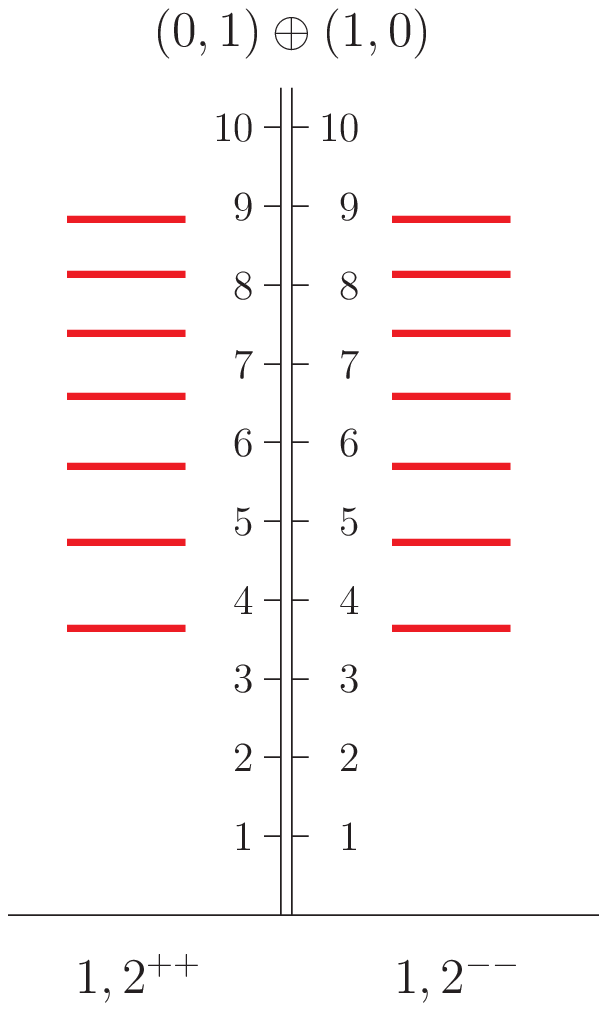}
}
\caption{ $\mu=0.2\sqrt{\sigma}$.
Spectra for chemical potential $\mu=0.2\sqrt{\sigma}$ . 
For $J=0$ only the two $(\frac{1}{2},\frac{1}{2})$ multiplets are 
present (left two panels). For $J>0$ there are also the 
$(0,0)$ and $(0,1)\oplus(1,0)$ multiplets. In the remaining four 
panels we show all multiplets for $J=2$. Masses are in units of $\sqrt \sigma$.
Meson quantum numbers are $I,J^{PC}$.
}
\end{figure*}

\medskip
{\bf 3. Conclusions.}  We have demonstrated
that it is possible to have a confining but chirally symmetric matter
consisting of chirally symmetric hadrons at finite density. 
Whether such a chirally symmetric but confining
phase exists in QCD or not is still an open question, 
but if it does, then it will imply dramatic modifications
of the QCD phase diagram.  It will also have
significant implications for astrophysics: The interactions of these
chirally-symmetric hadrons can be only of short-range as they decouple
from the Goldstone bosons and their  weak decay rate is quite different
since their axial  charge vanishes \cite{gl}.

\section*{Acknowledgments}
L. Ya. G. acknowledges suppoort of the Austrian Science
Fund through the grant P19168-N16.


\begin{thebibliography}{0}
\bibitem{pisarski} L. McLerran and R. D. Pisarski, {\it Nucl. Phys. A} {\bf 796},  
83 (2007).

\bibitem{GW} L. Ya. Glozman and R. F. Wagenbrunn, arXiv:0709.3080 [hep-ph].

\bibitem{Orsay}  A. Le Yaouanc, L. Oliver, O. Pene, and J. C.Raynal,
{\it Phys. Rev. D}  {\bf 29}, 1233 (1984); {\bf 31},  137 (1985).

\bibitem{Adler:1984ri}
S.~L. Adler and A.~C. Davis, {\it Nucl. Phys. B} {\bf 244}, 469 (1984).

\bibitem{Thies} V. Sch\"on and M. Thies, 
{\it Phys. Rev.} {\bf D 62}, 096002 (2000).


\bibitem{WG2} R. F. Wagenbrunn and L. Ya. Glozman,
{\it Phys. Rev. D} {\bf  75}, 036007 (2007).

\bibitem{GPR} L. Ya. Glozman, {\it Phys. Rep.} {\bf 444}, 1 (2007).

\bibitem{gl} L. Ya. Glozman, {\it Phys. Rev. Lett. } {\bf 99}, 191602 (2007).
\end{thebibliography}
\end{document}